\documentclass[aps,pra,showpacs,floatfix,twocolumn]{revtex4}

\usepackage{amsmath}
\usepackage{amsfonts}   
\usepackage{epsfig}

\newcommand{\ui}{\boldsymbol{\hat \imath}}
\newcommand{\uj}{\boldsymbol{\hat \jmath}}
\newcommand{\uk}{\boldsymbol{\hat k}}

\newcommand{\newvec}[1]{\boldsymbol{#1}}

\begin{document}

\begin{abstract}
It is well-known that an electric charge under a uniform magnetic field has a 
bidimensional motion if its initial position and velocity are perpendicular to 
this magnetic field. Although some constants of motion, as the energy and 
angular momentum, have been identified for this system, its features hide 
others. In this work, we build generalizations of the angular momentum and the 
Laplace-Runge-Lenz vector and show that these vectors are constants of motion.
Moreover, from them, we find four dynamically independent conserved qualities.
\end{abstract}

\title{Constants of motion for the magnetic force: the angular momentum and 
the Laplace-Runge-Lenz vector}
\author{D. Velasco-Mart\'inez, V. G. Ibarra-Sierra, J. C. Sandoval-Santana, 
A. Kunold and  J.L. Cardoso}
\affiliation{
   \'Area de F\'isica Te\'orica y Materia Condensada,
   Universidad Aut\'onoma Metropolitana at Azcapotzalco,
   Av. San Pablo 180, Col. Reynosa-Tamaulipas, Azcapotzalco,
   02200 M\'exico D.F., M\'exico } 

\maketitle

\section{Introduction}

An important approach to understanding and studying a classical mechanics
system is to identify its constants of the motion (CsM). There are several
methods to find them for a given system. The methods that rely on a systematic
and direct calculation are often efficient in finding the CsM but leave
the open question as whether there are other CsM \cite{LDLandau1, HGoldstein3, 
HIro}.

The dynamics of a charged particle in uniform electric and magnetic
field have been widely studied by diverse methods \cite{HGoldstein3, HIro,    
LDLandau2}. Many CsM as the energy \cite{HGoldstein3, HIro, LDLandau2}, 
angular momentum \cite{HIro, LDLandau2} and pseudomomentum \cite{HIro}
have been identified for this system.

The concept of the Laplace-Runge-Lenz vector (LRLV) has been used to describe 
the motion of two bodies interacting by a central $k/r$ potential 
\cite{LDLandau1, HGoldstein3, HIro, PSLaplace, CRunge, WLenz, HGoldstein1, 
HGoldstein2}. The LRLV vector was also found for an isotropic harmonic 
oscillator with central potential $\kappa r^2/2$ \cite{YGrandati}. In 
both problems, the angular momentum, given only by its mechanical part 
($\newvec{L}_{mech} =  \newvec{r} \times m\dot{\newvec{r}}$), is conserved.
Even more, for the classical motion of two charged particles, confined to two 
dimensions and embedded in a constant magnetic field, a component of the LRLV 
was obtained \cite{GMunoz} even though in this case $\newvec{L}_{mech}$ is not 
conserved and the magnetic force is not a central one.
However, the LRLV has been overlooked as a constant of motion (CM) of the
system comprised of a charged particle in uniform and perpendicular electric 
and magnetic fields.

The aim of this work is to study some of the CsM of a charged particle in 
uniform electric and magnetic perpendicular fields. Particularly we introduce 
LRLV as a CM for such a system.

This work is organized as follows. In Sec. \ref{orbit:OE} we present the 
classical analysis of a charged particle in uniform electric and magnetic 
fields in order to introduce the main concepts and physical quantities. Here 
we also show that the angular momentum is a CM and, from it, we obtain the 
charged particle's trajectory. In Sec. \ref{conservedmotion} we calculate the
LRLV and show that it is a CM. By using the obtained LRLV we integrate charged 
particle's orbit. By means of the Liouville theorem, we demonstrate that the 
angular momentum, the LRLV and the pseudomomentum are in fact CsM. Moreover, 
we prove that they are four dynamically independent conserved qualities. In 
Sec. \ref{conclusions}, we summarize.

\section{The Lorentz force and the Newton's second Law. The orbit equation}
\label{orbit:OE}

In order to study the charged particle motion we introduce Lorentz force
into Newton's second law of motion
\begin{equation} \label{Lorentz:fields0}
\frac{d}{dt} \left( m \dot{\newvec{R}} \right) = q \left(\newvec{E} + 
\dot{\newvec{R}} \times \newvec{B} \right)
\end{equation}
where $m$ is the mass, $q$ is the charge, $\newvec{R}$ is the particle's 
position, and $\newvec{E}$ and $\newvec{B}$ are the uniform external electric 
and magnetic fields respectively. As a first step we move into the frame of 
the guiding center coordinates by defining $\newvec{R} = \newvec{r} + 
(E/B)t\newvec{u}$ where $E/B$ is the drift velocity and $\newvec{u}$ is a 
unitary vector perpendicular to $\newvec{E}$ and $\newvec{B}$ such that
$\newvec{u} = \newvec{B} \times \newvec{E}/(BE)$. In this frame Newton's second
law takes the form
\begin{equation} \label{Lorentz:fields}
\frac{d}{dt} \left( m \dot{\newvec{r}} \right) = q  
\dot{\newvec{r}} \times \newvec{B}.
\end{equation}
From the previous results it is clear that the electron orbits around the 
center guiding coordinates $(E/B)t\newvec{u}$, and it in turn moves at 
constant speed $E/B$ in a direction $\newvec{u}$ perpendicular to the electric 
and magnetic fields.

Let us choose the polar coordinates ($r,\theta$) to obtain the orbit equation.
The equations of motion are:
\begin{eqnarray}
\label{motion:r}
\ddot{r} - r\dot{\theta}^2 &=& r\dot{\theta} \omega,\\
\label{motion:theta}
2 \dot{r} \dot{\theta} +r \ddot{\theta} &=& -\omega \dot{r},
\end{eqnarray}
where $\omega = qB/m$ is the cyclotron 
frequency.
The last expression can be expressed as a total time derivative
in the following form
\begin{equation}  \nonumber
\frac{d}{dt} \left( m r^2 \dot{\theta} +\frac{1}{2} q r^2 B  \right) =0,
\end{equation} 
therefore, as is shown in the next section,
the component parallel to $\newvec{B}$ of the
angular momentum with respect to the guiding coordinates
\begin{equation} \label{angularmomentum}
L = m r^2 \dot{\theta} +\frac{1}{2} q r^2 B
\end{equation}
is a CM\cite{HIro,LDLandau2}.
 
By introducing the solution for $\dot{\theta}$ from Eq. (\ref{motion:theta}) 
into (\ref{motion:r}), we obtain
\begin{equation} \label{motion:Hooke}
\ddot{r} - \frac{L^2}{m^2 r^3} + \frac{1}{4} \omega^2 r = 0,
\end{equation}
that corresponds to the motion equation of an isotropic harmonic oscillator.
Thus, the particle's orbit with respect to the guiding coordinate frame is (see 
Appendix \ref{orbit:talacha})
\begin{equation} \label{orbit}
r^2 -2 r r_0 \cos \left( \theta - \theta_0 \right) +r_0^2 = r_0^2-
\frac{2L}{m\omega} = a^2,
\end{equation}
which is the polar equation for a circle with radius $a$ centered
in $(r_0\cos\theta_0,r_0\sin\theta_0)$.

\section{The angular momentum, the Laplace-Runge-Lenz vector and the 
pseudomomentum}
\label{conservedmotion}

Let us rewrite Newton's second law of motion (\ref{Lorentz:fields})
for a charged particle in the guiding coordinate frame
\begin{equation} \label{Lorentz:potentials}
\frac{d}{dt} \left( m \dot{\newvec{r}} +q \newvec{A} \right) = -\nabla \left( 
q \phi - q  \dot{\newvec{r}} \cdot \newvec{A} \right)-q\newvec{E}
\end{equation}
where $\phi$ is the scalar potential and $\newvec{A}$ is the vector potential 
\cite{HGoldstein3, LDLandau2} that follow the usual relations $\newvec{E} = 
-\nabla \phi-\partial \newvec{A}/\partial t$ and $\newvec{B} = \nabla \times 
\newvec{A}$. Notice that the left side of the previous equation corresponds to 
the time variation of the minimal momentum 
\begin{equation}
\newvec{P} = m \dot{\newvec{r}} +q \newvec{A}. 
\end{equation}
Now we choose a gauge such that $\phi =-\newvec{r}\cdot\newvec{E}$ with 
uniform electric and magnetic fields. By taking a cross product of 
(\ref{Lorentz:potentials}) with $\newvec{r}$ for the left side we have
\begin{multline}
\frac{d}{dt} \left[ \newvec{r} \times \left( m \dot{\newvec{r}} +q \newvec{A} 
\right) \right] = q \left[ \newvec{r} \times \nabla \left( \dot{\newvec{r}} 
\cdot \newvec{A} \right)  + \dot{\newvec{r}} \times \newvec{A} \right]
\end{multline}
For an uniform and a constant magnetic field $\newvec{B}$ the vector potential 
in the Landau gauge is $\newvec{A} = (1/2) \newvec{B} \times \newvec{r}$ 
By replacing the explicit form of the vector potential, we obtain
\begin{multline}
\frac{d}{dt} \left[ m \newvec{r} \times \dot{\newvec{r}} + \frac{1}{2} r^2 q
\newvec{B} \right] \\
= \frac{q}{2} \left[ \newvec{r} \times \nabla \left( 
\dot{\newvec{r}} \cdot \newvec{B} \times \newvec{r} \right) + 
\dot{\newvec{r}} \times \left( \newvec{B} \times \newvec{r} \right) \right] .
\end{multline}
It is straightforward to show that the bottom side of this equation vanishes
and therefore the angular momentum, $\newvec{L} = \newvec{r} \times 
\newvec{P}$, is a CM. The modulus of $\newvec{L}$ is given by 
(\ref{angularmomentum}) in polar coordinates and it has the same direction as 
the magnetic field, while Cartesian coordinates it is given by
\begin{equation} \label{angularmomentum:cartesian}
L = m \left( x\dot{y}-y\dot{x} \right) + \frac{1}{2} m\omega \left(x^2 +y^2 
\right).
\end{equation}

On the other hand, by taking the cross product of (\ref{Lorentz:fields}) with 
$\newvec{L}$ for the right side, we have
\begin{equation}
\frac{d}{dt} \left(m \dot{\newvec{r}} \right) \times \newvec{L} = q \left( 
\dot{\newvec{r}} \times \newvec{B} \right) \times \newvec{L}.
\end{equation}
By using the facts that $\newvec{L}$ is a CM, $(d/dt) \left( m 
\dot{\newvec{r}} \right) \times \newvec{L} = (d/dt) \left(m 
\dot{\newvec{r}} \times \newvec{L} \right) $ and that $\newvec{L}$ is 
perpendicular to $\dot{\newvec{r}}$ we obtain
\begin{equation}
\left( \dot{\newvec{r}} \times 
\newvec{B} \right) \times \newvec{L} = - (d/dt) \left( BL \newvec{r} 
\right).
\end{equation}
We thus obtain a vectorial CM given by
\begin{equation} \label{LRLvector}
\newvec{T} = m \dot{\newvec{r}} \times \newvec{L} +qBL \newvec{r}.
\end{equation}
Given that we followed a similar method to the one used to calculate
the LRLV in the Kepler problem, we name it
the LRLV. It is perpendicular to the angular momentum and the magnetic field 
thus $\newvec{L} \cdot \newvec{T} =0$ and $\newvec{B} \cdot \newvec{T} =0$.

Since the LRLV is a CM, it can used as a basis to integrate the trajectory 
of the electron. As a first step we express $\newvec{T}$ in Cartesian 
coordinates as
\begin{equation} \label{LRLvector:cartesian}
\newvec{T} = T_x \ui +T_y \uj,
\end{equation}
where
\begin{equation} \label{LRLV:x} 
T_x = Lm \left( \dot{y} +\omega x \right)
\end{equation}
 and 
\begin{equation} \label{LRLV:y} 
T_y = -Lm \left( \dot{x} -\omega y \right)
\end{equation}
are the Cartesian components of the LRLV. Second, we obtain the 
modulus of the LRLV as 
\begin{equation} \label{LRLvector:mod}
T = L \sqrt{2m \left( E +L\omega \right)}.
\end{equation}  
Finally, the equation for the particle's trajectory is obtained by solving 
$\dot{x}$ and $\dot{y}$ from (\ref{LRLV:x}) and (\ref{LRLV:y}) respectively and 
substituting the result  in (\ref{angularmomentum:cartesian}) giving
\begin{equation} \label{LRLV:orbit}
\left( x -\frac{T_x}{Lm\omega} \right)^2 + \left( y -\frac{T_y}{Lm\omega} 
\right)^2 = \frac{2E}{m \omega^2}.
\end{equation}
This expression is the circle equation with center at $\left( T_x/Lm\omega, 
T_y/Lm\omega \right)$ and radius $a = \sqrt{2E/m\omega^2}$, where the 
mechanical energy is associated with the initial momentum ($E = (1/2m)
\left(p_{x0}^2 +p_{y0}^2 \right)$). In polar coordinates we obtain an orbit 
equation which is identical to Eq. (\ref{orbit}), if we set $\tan \theta_0 = 
T_y / T_x$ and $r_0 = T/m\omega L$. Notice that the LRLV is parallel to the 
center of the trajectory, in fact
\begin{equation}
\newvec{r}_0 = \frac{\newvec{T}}{m \omega L} = \left( \frac{\dot{y}}{\omega} + 
x \right) \ui -\left( \frac{\dot{x}}{\omega} - y \right) \uj.
\end{equation}
For a first guess as to the direction of the vector $\newvec{r}_0$ it is 
helpful to compute $\newvec{r}_0 \cdot \newvec{L}$. Because of the 
orthogonality of $\newvec{L}$ to both terms in the definition of 
$\newvec{r}_0$ this dot product vanishes.  From this result it follows that 
$\newvec{r}_0$ must lie in the particle's orbit plane of motion. As we 
calculated above, this interpretation of the LRLV implies that $\newvec{r}_0$ 
should be conserved because the position and geometry of a bound orbit does 
not change over time and therefore it should depend on the initial conditions 
e. g. the particle's initial position and velocity. Let us set $T_x/Lm 
= \omega x_0$ and $T_y/Lm = \omega y_0$ where  $x_0 = r_0 \cos \theta_0$ and 
$y_0 = r_0 \sin \theta_0$ and Eqs. (\ref{LRLV:x}) and (\ref{LRLV:y}) describe 
the movement of the charge particle. 

According to Ref. \cite{HIro}, the last conserved vector can be obtained 
by rewriting Eq. (\ref{Lorentz:fields}) as
\[
\frac{d}{dt} \left[ m \dot{\newvec{r}} -q \newvec{r} \times \newvec{B} \right] 
= 0.
\]
In this way, we have a third conserved vector
\begin{equation} \label{pseudomomentum}
\newvec{P}_s =  m \dot{\newvec{r}} -q \newvec{r} \times \newvec{B} ,
\end{equation}
this vector is {\it not} the minimal momentum. From now on, we name as 
{\it the pseudomomentum}. In Cartesian coordinates, this vector is given by
\begin{equation} \label{pseudomomentum:cartesian}
\newvec{P}_s =  m \left( \dot{x} -\omega y\right) \ui + m\left( \dot{y} +\omega 
x\right) \uj.
\end{equation} 
Notice that $\newvec{P}_s \cdot \newvec{B} =0$, $\newvec{P}_s \cdot \newvec{L} 
=0$ and $\newvec{P}_s \cdot \newvec{T} =0$, by taking the cross product with 
$\newvec{r}_0$, we get
\begin{equation}
\newvec{r}_0 \times \newvec{P}_s = \frac{2}{\omega} \left( E +L\omega \right) 
\uk 
\end{equation}
and its modulus is
\begin{equation} \label{pseudomomentum:modulus}
P_s = \sqrt{ 2m \left(E+L\omega \right)} .
\end{equation} 
The LRLV is also proportional to $E+L\omega$ as can be seen in Eq. 
(\ref{LRLvector:mod}), moreover $T = P_s L$. It can be shown that this 
quantity is a CM \cite{HIro}. 

The Liouville Theorem is a well-known approach to test whether or not a 
quality is conserved. By defining the Poisson bracket in such way that
\begin{eqnarray} \label{Poisson}
\left\{ r, p_r \right\} &=& 1 \\
\left\{ \theta, p_\theta \right\} &=& 1
\end{eqnarray}
where $p_r = m \dot{r}$ and $p_\theta = L = mr^2 \left( \dot{\theta} +\omega/2 
\right)$, we can find that the vectors $\newvec{L}$, $\newvec{T}$ and 
$\newvec{P}_s$ are CsM. With those three vectors, the system has ten CsM: 
the energy, the three components of the angular momentum, the three 
components of the LRLV and the three components of the pseudomomentum. Because 
this system has six initial conditions, the three components of the position 
vector and the three components of the initial momentum, there must exist four 
relations that turn these ten dependent CsM into six independent ones, namely, 
there must be four relations connecting these qualities. Such relations are the 
orthogonality of $\newvec{B}$, $\newvec{L}$, $\newvec{T}$ and $\newvec{P}_s$, 
i. e. $\newvec{L} \cdot \newvec{T} =0$, $\newvec{L} \cdot \newvec{P}_s =0$, 
$\newvec{P}_s \cdot \newvec{T} =0$, $\newvec{B} \cdot \newvec{T} =0$ y 
$\newvec{B} \cdot \newvec{P}_s =0$. The first and second relations imply that 
$L$, Eq. (\ref{angularmomentum}), is a CM; $\newvec{P}_s \cdot \newvec{T} =0$ 
indicates that $E +L \omega$ is another CM; and, $\newvec{B} \cdot \newvec{T} 
=0$ and  $\newvec{B} \cdot \newvec{P}_s =0$ show that $\omega$ is also a CM. 
Finally, the four relations  
\begin{eqnarray}
\omega &=& \frac{qB}{m} \\
L &=& mr^2 \left( \dot{\theta} +\frac{\omega}{2} \right) \\
E &=& \frac{p_r^2}{2m} +\frac{L^2}{2mr^2} -\frac{L \omega}{2} 
+\frac{1}{8} m \omega^2 r^2 \label{energy} \\
\frac{T^2}{2mL^2} &=& E +L \omega . \label{LRLV:mod2}
\end{eqnarray}
are dynamically independent, because they are in involution, i. e. 
$\left\{ C_i , C_j \right\} = 0$ where $C_i = \omega, L, E {\rm \ or \ } 
T^2/2mL^2$. Those relations, after some rearranging, give the nine dependent 
components of $\newvec{L}$, $\newvec{T}$ and $\newvec{P}_s$ in 
terms of the cyclotron frequency, the energy, the angular momentum and 
$T^2/2mL^2$.

\section{Conclusions}
\label{conclusions}

We have studied the CsM of a charged particle
in uniform electric and magnetic fields. Aside from the
well known CsM as the energy\cite{LDLandau1, HGoldstein3, 
HIro},
we found that a vector, obtained by similar means as the LRLV, is also a
CM connected to the center of the particle's orbit.
The particle's trajectory was integrated from it.
Additionally we have proved that
the cyclotron frequency $\omega$, the angular momentum $L$, the
energy $E$ and the LRLV $T$ are four dynamically conserved qualities.

\appendix

\section{The particle's trajectory}
\label{orbit:talacha}

\begin{figure}
\includegraphics[angle=0,width=3.2in]{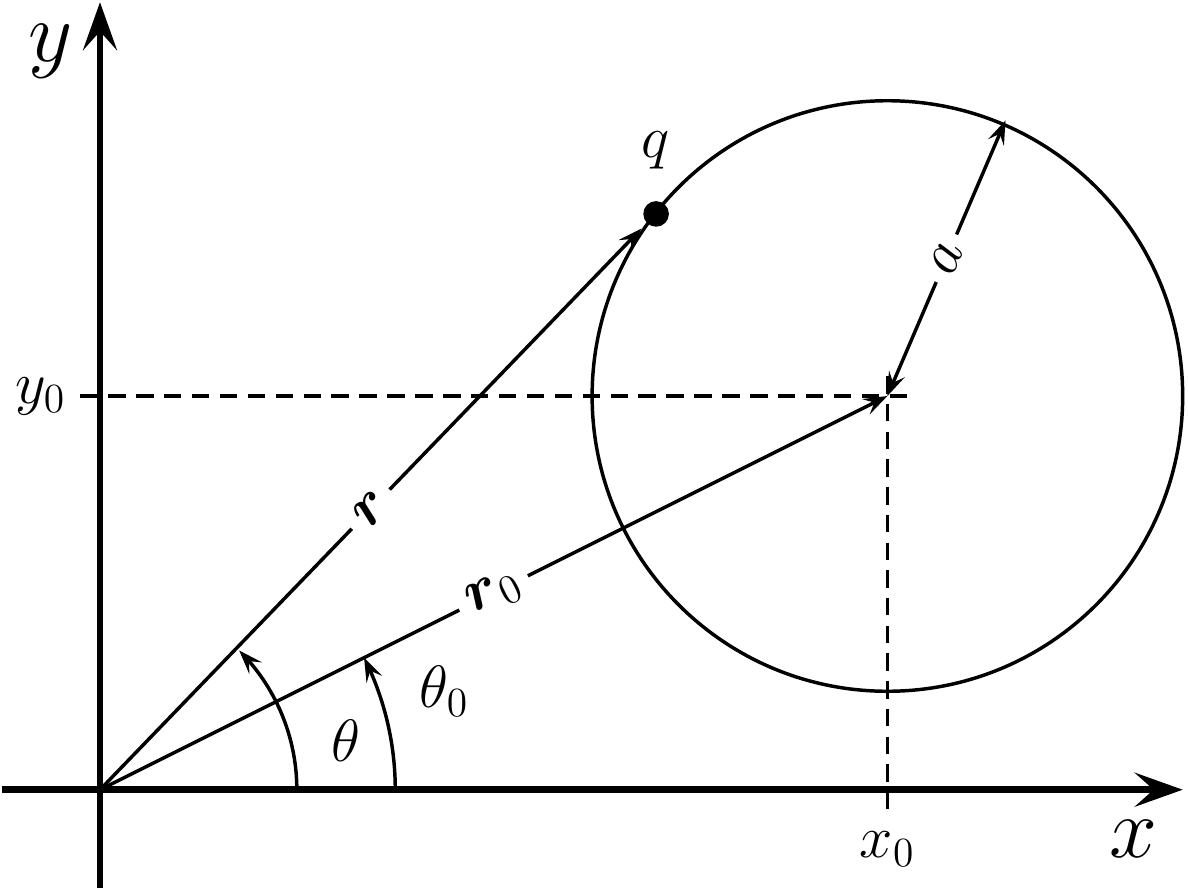}
\caption{Eq. $r^2 -2 r r_0 \cos \left( \theta -\theta_0 \right) +2L/m\omega 
= 0$ is plotted in the plane $xy$ and it describes a circle of radius $a = 
\sqrt{r_0^2 -2L/m\omega}$ and centered at $\left( x_0, y_0 \right)$ where $x_0 
= r_0 \cos \theta_0$ and $y_0 = r_0 \sin \theta_0$.}
\label{plot}
\end{figure}

Here we obtain the particle's trajectory by using the angular momentum
and later on the LRLV.
We start by expressing $r$ as a function of $\theta$ and performing
the variable change $r=u^{-2}$
in the differential equation
(\ref{motion:Hooke})
\begin{eqnarray}
\nonumber \frac{dr}{dt} &=& \frac{dr}{du} \frac{du}{d \theta} \frac{d 
\theta}{dt} \\
\nonumber \frac{dr}{dt} &=& -\left( \frac{L}{m} \frac{d}{d \theta} u^2 + 
\frac{1}{2} \omega \frac{d}{d \theta} u^{-2} \right).
\end{eqnarray}
Doing a new variable change
\[
g = \frac{L}{m} u^2 + \frac{1}{2} \omega u^{-2}
\]
we can obtain $\ddot{r}$
\[
\ddot{r} = -\left( \frac{L}{m} u^4 - \frac{1}{2} \omega \right) \frac{d^2 g}{d 
\theta^2}
\]
On the right hand side of the previous
equation, the two last terms can be 
factorized as 
\[
-\frac{L^2}{m^2} u^6 + \frac{1}{4} \omega u^{-2} = -\left( \frac{L}{m} u^4 - 
\frac{1}{2} \omega \right) g.
\]
With the two previous expressions, we can write down Eq. (\ref{motion:Hooke})  
in the following form
\begin{equation}
\frac{d^2 g}{d \theta^2} +g =0.
\end{equation}
The solution of the previous differential equation yields
\begin{equation} \label{orbit:doe}
r^2 -2 r r_0 \cos \left( \theta -\theta_0 \right) +\frac{2L}{m \omega} = 0,
\end{equation}
In Cartesian coordinates, this 
expression can be rewritten as
\begin{equation}
\left( x-x_0 \right)^2 +\left( y-y_0 \right)^2 = a^2.
\end{equation}
It describes a circular trajectory with radius $a = \sqrt{r_0^2 -2L/m\omega}$ and center 
at $\left( x_0, y_0 \right)$ where $x_0 = r_0 \cos \theta_0$ and $y_0 = r_0 
\sin \theta_0$ as shown in Fig. \ref{plot}. 

Similarly, by solving $\dot{r}$ from (\ref{energy}), using (\ref{LRLV:mod2})
and applying the 
previous variable change  we obtain
\[
\left( \frac{dg}{d\theta} \right)^2 = \frac{2}{m} \left(E +L\omega\right) -g^2 
= \frac{T^2}{m^2 L^2} -g^2,
\]
Finally we integrate and obtain the 
orbit expressed in Eq. (\ref{orbit:doe}).

\bibliography{biblio}

\end{document}